\documentclass[a4paper,twosides]{article}
\usepackage{multicol,multirow,graphics,fancyhdr,epstopdf}  
\usepackage{amsmath,amsfonts,amssymb,bm,upgreek,mathrsfs,mathcomp}  
\usepackage[pagewise,switch,columnwise]{lineno}
\usepackage[compress,nospace]{cite}
\usepackage{CPL-2022}
\usepackage{multicol}
\usepackage{ulem}
\usepackage{graphicx}
\usepackage{caption}
\usepackage[catcode]{changes}  

\newcommand{\software}[1]{\texttt{#1}}

 \newcommand{\cplvol}{xx}
\newcommand{\cplno}{x} \newcommand{\cplpagenumber}{xxxxxx}
\newcommand{\blue}[1]{\textcolor{blue}{{#1}}}

\begin{document}
\begin{center}
\large\bf{\boldmath{A narrowband burst from FRB 20190520B simultaneously observed by FAST and Parkes}}
\footnotetext{\hspace*{-5.4mm}$^{*}$Corresponding authors. Email: niuchenhui@ccnu.edu.cn; dili@nao.cas.cn

}
\\[5mm]
\normalsize \rm{}Yuhao Zhu$^{1,2}$, Chenhui Niu$^{3*}$, Shi Dai$^{4}$, Di Li$^{5,1*}$, Pei Wang$^{1,6}$, Yi Feng$^{7,8}$, Jingwen Wu$^{2,1}$,\\ Yongkun Zhang$^{1,2}$, Xianghan Cui$^{1,2,9}$, Junshuo Zhang$^{1,2}$, Jinhuang Cao$^{1,2}$
\\[3mm]\small\sl $^{1}$National Astronomical Observatories, Chinese Academy of Sciences, Beijing 100101, China

$^{2}$University of Chinese Academy of Sciences, Beijing 100049, China

$^{3}$Institute of Astrophysics, Central China Normal University, Wuhan 430079, China

$^{4}$Western Sydney University, Locked Bag 1797, Penrith South DC, NSW 2751, Australia

$^{5}$Department of Astronomy, Tsinghua University, Beijing 100084, China

$^{6}$Institute for Frontiers in Astronomy and Astrophysics, Beijing Normal University, Beijing 102206, China

$^{7}$Research Center for Astronomical Computing, Zhejiang Laboratory, Hangzhou 311100, China

$^{8}$Institute for Astronomy, School of Physics, Zhejiang University, Hangzhou 310027, China

$^{9}$International Centre for Radio Astronomy Research, Curtin Institute of Radio Astronomy, Perth 6102, Australia
\\[4mm]\normalsize\rm{}(Received xxx; accepted manuscript online xxx)
\end{center}
\vskip 1.5mm

\small{\narrower 
Fast Radio Bursts (FRBs) are short-duration radio transients with mysterious origins.
Since its uncertainty, there are very few FRBs that are observed by different instruments, simultaneously.
This study presents a detailed analysis of a burst from FRB 20190520B observed by FAST and Parkes at the same time. 
The spectrum of this individual burst ended at the upper limit of the FAST frequency band and was simultaneously detected by the Parkes telescope in the 1.5-1.8 GHz range.
By employing spectral energy distribution (SED) and spectral sharpness methods, we confirmed the presence of narrowband radiation in FRB 20190520B, which is crucial for understanding its radiation mechanisms.
Our findings support the narrowband characteristics that most repeaters exhibit. 
This work also highlights the necessity of continued multiband observations to explore its periodicity and frequency-dependent properties, contributing to an in-depth understanding of FRB phenomena.
\par}\vskip 3mm
\vskip 5mm

\begin{multicols}{2}
Fast radio bursts (FRBs) are bright, transient astronomical events lasting from microseconds to milliseconds, observed in radio wavelengths.\ucite{1} 
Their high dispersion measures (DMs), which exceed the contributions from our galaxy, indicate an extragalactic origin.\ucite{2,3}
The field of FRB research has undergone a rapid development in terms of observational analysis, leading to significant progress in understanding this enigmatic astronomical event.\ucite{4,5,6,7}
Numerous models have been proposed to explain the origins, radiation mechanisms, and plausible progenitors of FRBs.
Platts et al. presented an overview of the radiation mechanisms and potential progenitors of FRBs.\ucite{8}
Zhang (2020) systematically summarized and outlined a variety of theoretical models that could explain the origins of FRBs.\ucite{9}
Li et al. suggested that magnetars are probable progenitors of some FRBs, although it remains uncertain whether all FRBs originate from magnetars.\ucite{10}

In these detected FRBs, some are observed with only one burst and ceased after, which are called apparent non-repeating FRBs, or one-offs, while some are detected with repeated bursts, known as repeating FRBs, or repeaters.
Previous research indicates that bursts from repeating sources, such as FRB 20121102A, generally last longer and have a narrower spectrum compared to those from non-repeating sources. 
Additionally, a downward frequency drift was also noted as a common feature by Pleunis et al. in these repeaters.\ucite{11}
Numerous studies have explored the cosmological histories and redshift evolution of FRBs, revealing both distinct and overlapping patterns across different samples and classifications.\ucite{12,13,14,15,16,17}
Despite these findings, it remains uncertain whether all FRBs will eventually be observed to repeat if monitored over a sufficiently long period.

Considering that repeaters generate series of bursts that last for a period of time, monitoring these repeaters can significantly enlarge the burst sample, thereby facilitating more robust model interpretations, propagation effect studies, and other systematic investigations.
In 2019, the repeating fast radio burst, FRB 20190520B was initially detected by the Five-hundred-meter Aperture Spherical Telescope (FAST).
This discovery was made during the Commensal Radio Astronomy FAST Survey (CRAFTS) in drift-scan mode, within the frequency range of 1.05 to 1.45 GHz.\ucite{18,19,20}
Following the discovery of this source, VLA observations localized its host galaxy and further confirmed that a persistent radio source (PRS) was physically connected with FRB 20190520B, similar to the findings for FRB 20121102A.\ucite{20}
The DM of FRB 20190520B is $1204.70 \pm 4.0\, \mathrm{pc\, cm^{-3}}$, and subsequent VLA observations confirmed a corresponding host galaxy redshift of $z = 0.241 \pm 0.001$.\ucite{20}
Extensive research has been conducted on FRB 20190520B's source environment,\ucite{21,22} associated PRS,\ucite{23,24,25} polarization characteristics of repeating bursts,\ucite{26} and host galaxy.\ucite{27}
Additionally, numerous models have been proposed to explain the radiation mechanisms of this repeater.\ucite{28,29}

Multiband observations are crucial for understanding the nature and frequency-dependent properties of FRB radiation across different frequency bands, thereby refining our understanding of their radiation mechanisms.\ucite{30,31,32,33}
Feng et al. identified a trend of lower polarization at lower frequencies in repeating FRBs, modeled by multipath scattering, suggesting a complex environment near repeating FRBs.\ucite{34}
This provides key insights into the origins of repeaters.
Observations on repeater FRB 20180916A demonstrated a frequency-dependent active window, highlighting significant variability in FRB emissions across different frequencies.\ucite{30,35,36,49}
These observations help further clarify the radiation mechanisms of FRBs, improving our understanding of their nature across different bands.

Good et al. used data from CHIME and Arecibo to study FRB event rates, noting that Arecibo's non-detection of certain bursts observed by CHIME suggests the presence of unique characteristic frequencies for specific repeaters.\ucite{37}
In studying FRB 20180301A, Kumar et al. discovered that bursts above 1.8 GHz were rare, with the majority occurring around 1.1 GHz, suggesting narrowband characteristics with a spectral coverage $\Delta \nu / \nu_c \sim 0.05-0.83$.\ucite{38}
Similarly, Zhang et al.'s analysis of FRB 20220912A using FAST observations revealed a spectral coverage between 0.1 and 0.4.
These findings underscore the necessity for detailed spectral analysis in FRB research.\ucite{39}
The lack of multi-band observation data makes it challenging to analyze the energy evolution, event rates, and the active window across frequencies.
Thus, comprehensive multi-band observations are imperative to determine the spectral limitations, radiation mechanisms, and origins of FRB 20190520B.

In this Letter, we first outline the observations conducted using the FAST and Parkes telescopes, followed by explaining the method employed to optimize the dispersion measure.
We analyze the data focusing on flux calibration, spectral energy distribution (SED), and spectral sharpness and spectral coverage of a particular burst from FRB 20190520B to study its narrowband emission features, which are common in repeaters.
We then discuss the SED fitting method used by CHIME et al.,\ucite{40} comparing it with the spectral coverage and our sharpness methods.
Finally, we summarize the conclusions of this study.

{\it Observation.}
In this work, FRB 20190520B was observed using two telescope facilities, FAST and Parkes. 
The observation campaign commenced with Parkes on August 6, 2020, and FAST began its observation epoch on October 24, 2020, with both telescopes conducting observations independently.
Figure~\ref{fig:1} shows the observation epochs.
After careful data reduction, we found that only one burst was simultaneously detected by both telescopes during this observation epoch.

{\it FAST.}
FAST provides high sensitivity for detecting weak pulses within the 1.05-1.45 GHz range.
During this epoch, observations of FRB 20190520B were carried out using the FAST telescope from October 24, 2020, to April 1, 2023.
The central beam of FAST's L-band 19-beam receiver was employed to search for pulses.
Data were recorded using the Pulsar backend in the PSRFITS format, with 8-bit quantization, a time sampling resolution of 98.304 $\mu s$, and a frequency resolution of 0.122 MHz across 4096 channels, covering a frequency range of 1000-1500 MHz.
Four polarization products are captured during each observation.

A noise diode signal with a temperature $T_\mathrm{noise} \simeq 12.5 \mathrm{K}$ will be periodically injected into the system with a 2.01326 second period at the leading 1-minute for calibration purposes before each Pulsar/FRB observation session in FAST.\ucite{41}
The injected noise calibration signal was used to scale the data into antenna temperature, measured in units of kelvin.
Then, according to the zenith-angle-dependent gain curve given by Jiang et al., the calibrated peak flux density and corresponding uncertainties are obtained.\ucite{41}
Frequencies outside the 1.05-1.45 GHz range were excluded due to gain defects.
Additionally, channels affected by significant radio frequency interference (RFI) were removed prior to further flux calibration and analysis.
\vskip 4mm
\fl{1}\centerline{\includegraphics[width=\columnwidth]{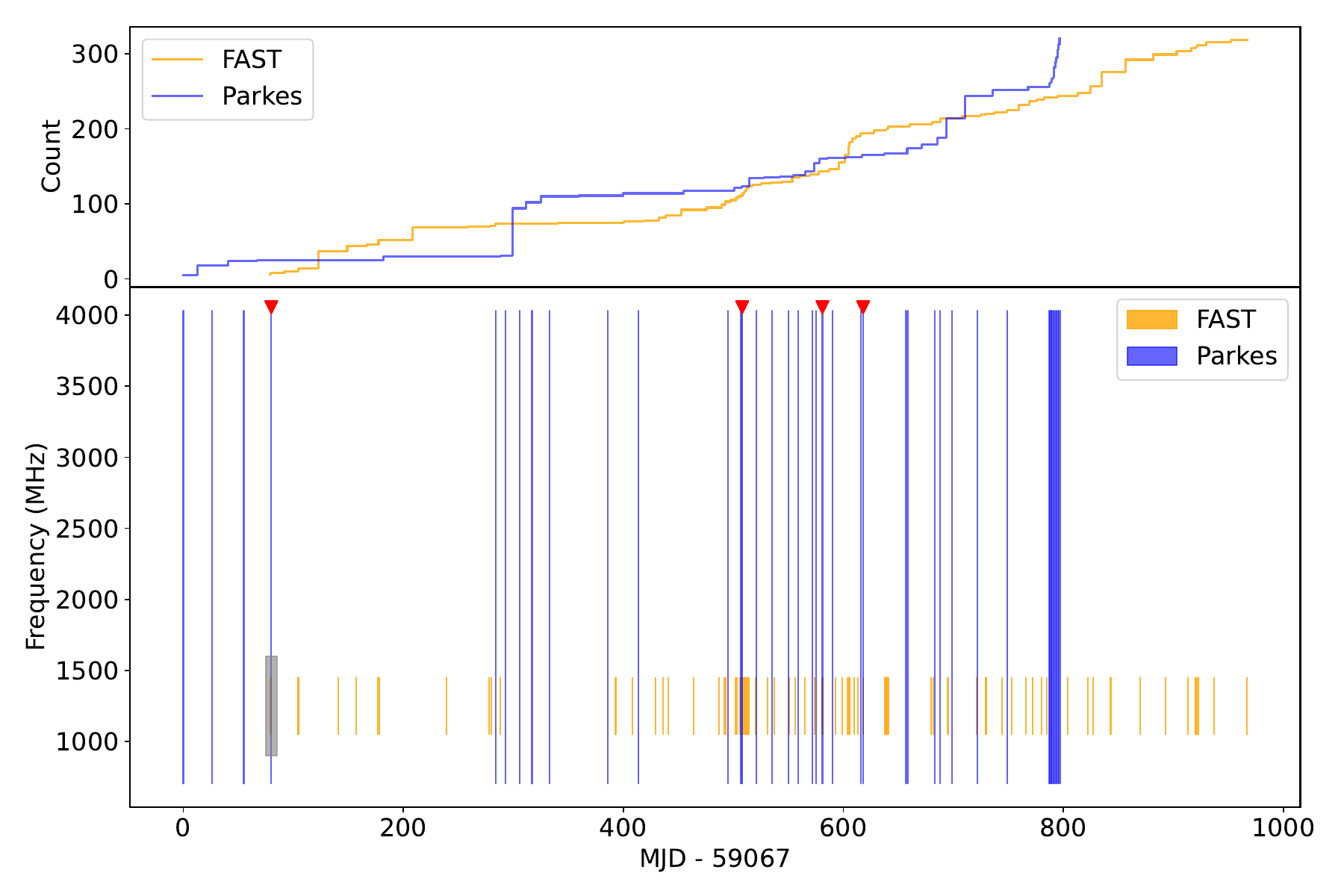}}
\vskip 2mm
\figcaption{7.5}{1}{
The observation epochs (in MJD) and the frequency coverage for FAST (orange) and Parkes (blue) are displayed.
The lower panel illustrates the bursts detected during multiband observations with lines, while the upper panel shows the cumulative burst count for both telescopes.
In this observation, most bursts were detected independently by the two telescopes due to minimal overlap in observation times. 
Joint detections are marked by red arrows, with the gray-shaded area indicating the presence of the simultaneously detected burst.
}
\medskip

We searched the FAST data using the GPU-based \software{heimdall} package, keeping candidates with $\mathrm{S/N} > 7$ for manual inspection.\ucite{42}
During this period, we conducted 73 observations and detected 318 bursts.
A detailed analysis of long-term FAST monitoring will be presented in a subsequent paper.

{\it Parkes.}
The Parkes 64-m diameter telescope is updated with an ultra-wide-bandwidth low-frequency receiver (UWL) in 2018. 
With this updated receiver, Parkes now covers a wide frequency range of 704-4032 MHz.\ucite{43}
The frequency coverage is wider than FAST, so it can detect more intact spectral energy distributions (SEDs) from pulses, expanding the understanding of radiation mechanisms of FRB 20190520B.

Parkes has been monitoring this source fortnightly from August 6, 2020, to October 12, 2022.
For repeaters with a known dispersion measure, which is the scenario in this study, the search-mode data stream was coherently de-dispersed to produce PSRFITS search-mode data files.
The data was recorded with a 32 $\mu s$ time sampling resolution, 8-bit samples, four polarization products, and 128 channels for each subband.
Subsequently, these files will be searched to identify and further verify candidates with $\mathrm{S/N} > 7$.
During this 41-day Parkes observation epoch, 320 bursts were detected.
Before each observation, square-pulse noise signals were injected for calibration. 
After calibration, flux density measurements were taken and the data was further cleaned to eliminate RFI using the \software{pazi} routine from \software{PSRCHIVE}, with careful flagging of problematic channels.\ucite{44}
In this Letter, detailed analysis of Parkes data will not be analyzed as a comprehensive paper by Dai et al. is currently under preparation.
\vskip 4mm
\fl{2}\centerline{\includegraphics[width=0.85\columnwidth]{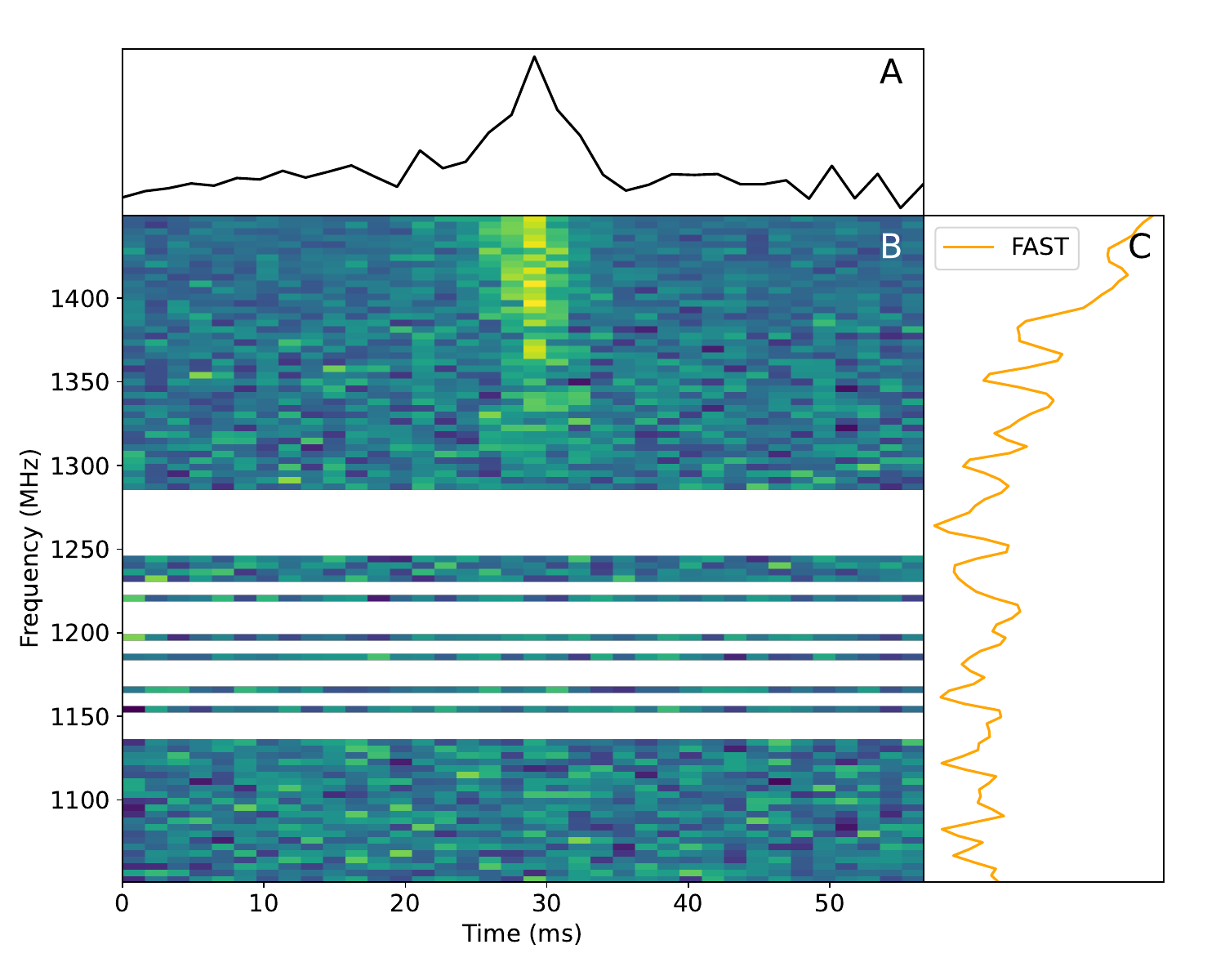}}
\vskip 2mm
\figcaption{7.5}{2}{
Panel B presents a waterfall plot of the burst detected by FAST between 1.05-1.45 GHz, with RFI channels appearing as white blanks.
Panel A shows the frequency-averaged profile, and Panel C illustrates the SED, indicating that there are no clear emission features below approximately 1200 MHz.
}
\medskip

\vskip 4mm
\fl{3}\centerline{\includegraphics[width=0.85\columnwidth]{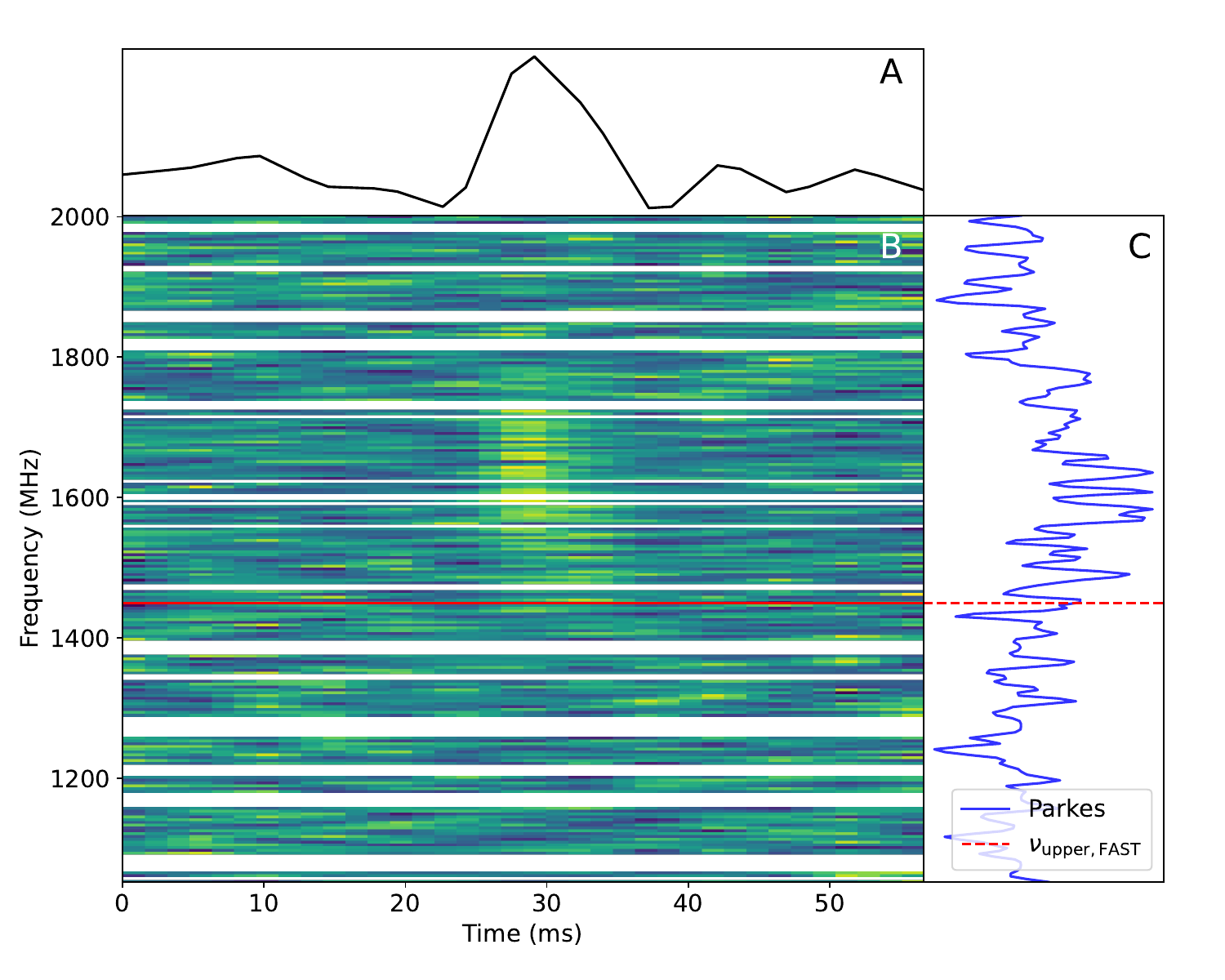}}
\vskip 2mm
\figcaption{7.5}{3}{
Panel B displays the waterfall plot of the burst detected by Parkes, originally covering the full bandwidth. 
For better visualization, the frequency range has been narrowed to 1.05-2.00 GHz.
}
\medskip

{\it Flux calibration.}
For each burst, we estimate the peak flux density using the radiometer equation
\begin{equation}
    S_{\nu} = \frac{\mathrm{SNR}\times T_{\mathrm{sys}}\times \beta}{G\times \sqrt{n_p\times \Delta \nu \times t_{\mathrm{samp}}}},
    \label{eq:1}
\end{equation}
where $T_{\mathrm{sys}}$ is the system temperature, $G$ is the gain of the telescope, $n_p=2$ is the polarisation channels, $\Delta \nu$ is the bandwidth, $\beta$ is the digitization factor, and $t_{\mathrm{samp}}$ is the sampling time of the burst profile.

Ideally, FAST has a system temperature \(T_{\mathrm{sys}} = 20 \, \mathrm{K}\) and a gain of \(G = 16.5 \, \mathrm{K \, Jy^{-1}}\).\ucite{41}
Considering Parkes, with a system temperature \(T_{\mathrm{sys}} = 22\sim 23 \, \mathrm{K}\), an aperture efficiency \(\epsilon_{\mathrm{ap}}=0.60\sim 0.67\), and an effective area \(A_e = 3217 \, \mathrm{m^2}\) at frequencies between 1088 and 1600 MHz, the calculated gain ranges from \(0.70\sim 0.82 \, \mathrm{K \, Jy^{-1}}\).\ucite{43}

Assuming that a burst detected by FAST is downsampled to match Parkes' frequency resolution of 4.0 MHz and time resolution of \(12.80\, \mathrm{\mu s}\), the flux density detection threshold for Parkes is \(\delta S_{\mathrm{Parkes}} = 0.84\sim 1.03 \, \mathrm{Jy}\), whereas for FAST, it is \(\delta S_{\mathrm{FAST}} = 0.03788 \, \mathrm{Jy}\). Consequently, FAST's sensitivity within its frequency band is approximately 22 to 27 times higher than that of Parkes.
 
{\it Analysis.}
During the multiband observation campaign, approximately 300 bursts were detected by each of the two telescopes, and of these bursts, one was observed simultaneously by both FAST and Parkes.
Figure~\ref{fig:2},\ref{fig:3} illustrates the burst simultaneously captured with Parkes.
This section focuses on analyzing the SED of a single burst detected simultaneously by both FAST and Parkes, to confirm the band-limited emission feature in FRB 20190520B.


{\it DM optimization.}
Precisely calculating the dispersion measure (DM) of the bursts is essential to examine the propagation of the radiation.
An inaccurate DM can obscure variations in the circum-source environment, leading to incorrect estimates of the radiation zone and its surroundings, thereby hindering our understanding of FRB radiation mechanisms.
We used the structure S/N maximum method with \software{DM\_phase.py} to determine the optimal DM of a burst.\ucite{50}
The algorithm generates a ``coherence spectrum'' for various DM values for the trial by applying a 1D Fourier transform to the intensity data in the frequency channels, normalizing by the amplitude and integrating over the emission bandwidth. 

For bursts with $\mathrm{S/N} > 7$, \software{DM\_phase.py} was applied to derive the corresponding DM, using a trial DM range of -10 to 10 $\mathrm{pc\, cm^{-3}}$ with 2000 steps.
The observed burst properties are shown in Table~\ref{tbl:1}.
\vskip 2mm
\tl{1}\tabtitle{7.8}{1}{
The burst properties of the simultaneously detected burst.
The barycentric MJD for this burst recorded by FAST and Parkes is 59147.270899223 and 59147.270899265, respectively.
}
\vskip 2mm \tabcolsep 4.5pt
\centerline{\footnotesize
\begin{tabular}{lccc}
\hline\hline\hline
& DM  & Width & Flux Density \\
& ($\mathrm{pc\, cm^{-3}}$)    & (ms)  &  (mJy)       \\
\hline
$\mathrm{Parkes}$    & $1200.81\pm 0.30$  & $5.53\pm 0.52$  & $1194\pm 160$   \\
$\mathrm{FAST}$      & $1198.49\pm 0.16$  & $5.76\pm 0.58$  & $53.23\pm 3.53$       \\
\hline\hline\hline
\end{tabular}}
\medskip
Because both telescopes observed the same burst, the dispersion measures (DMs) associated with the burst are consistent.
Given that the FAST telescope has higher sensitivity and resolution, which allows it to detect finer structures and provided a relatively higher flux for the burst, it offers better data for optimizing the DM. 
Consequently, we utilized the optimized DM derived from FAST data for subsequent analyses for this burst.

{\it Spectral energy distribution.}
The spectral energy distribution (SED) of FRB bursts represents the frequency response within the observed frequency band.
This may reveal the emission physics of FRBs.
First, we integrate the on-pulse region over time to obtain the SED, including background flux.
We then measure the off-pulse region to capture the background flux, which is subtracted from the on-pulse flux.
Following this, the spectrum is smoothed, and the burst bandwidth is fitted to determine the spectral coverage.
Finally, the flux is calibrated to finalize the SED and convert relative intensities into the flux density measured in Jansky (Jy).

We use the simultaneously detected burst to further confirm the emission feature of FRB 20190520B.
Parkes may miss weak emission components below its detection threshold, potentially leading to the erroneous conclusion that the emission follows a narrowband feature. 
In contrast, the higher sensitivity of FAST allows it to detect radiation components that Parkes might overlook.

Figure~\ref{fig:4} shows the SEDs for the burst detected simultaneously by FAST and Parkes.
The SED from Parkes, within the bandwidth of FAST, shows no distinct emission due to significant flux fluctuations, although an upward trend around 1500 MHz is possible.
The zoomed-in panel highlights FAST's higher sensitivity in detecting emission components below the threshold detectable by Parkes.
FAST's SED exhibits a band-limited feature, with no emission below 1200 MHz, indicating a narrowband and singular radiation component in the spectrum.
The rising trend at higher frequencies is consistent between FAST and Parkes.
\vskip 4mm
\fl{4}\centerline{\includegraphics[width=\columnwidth]{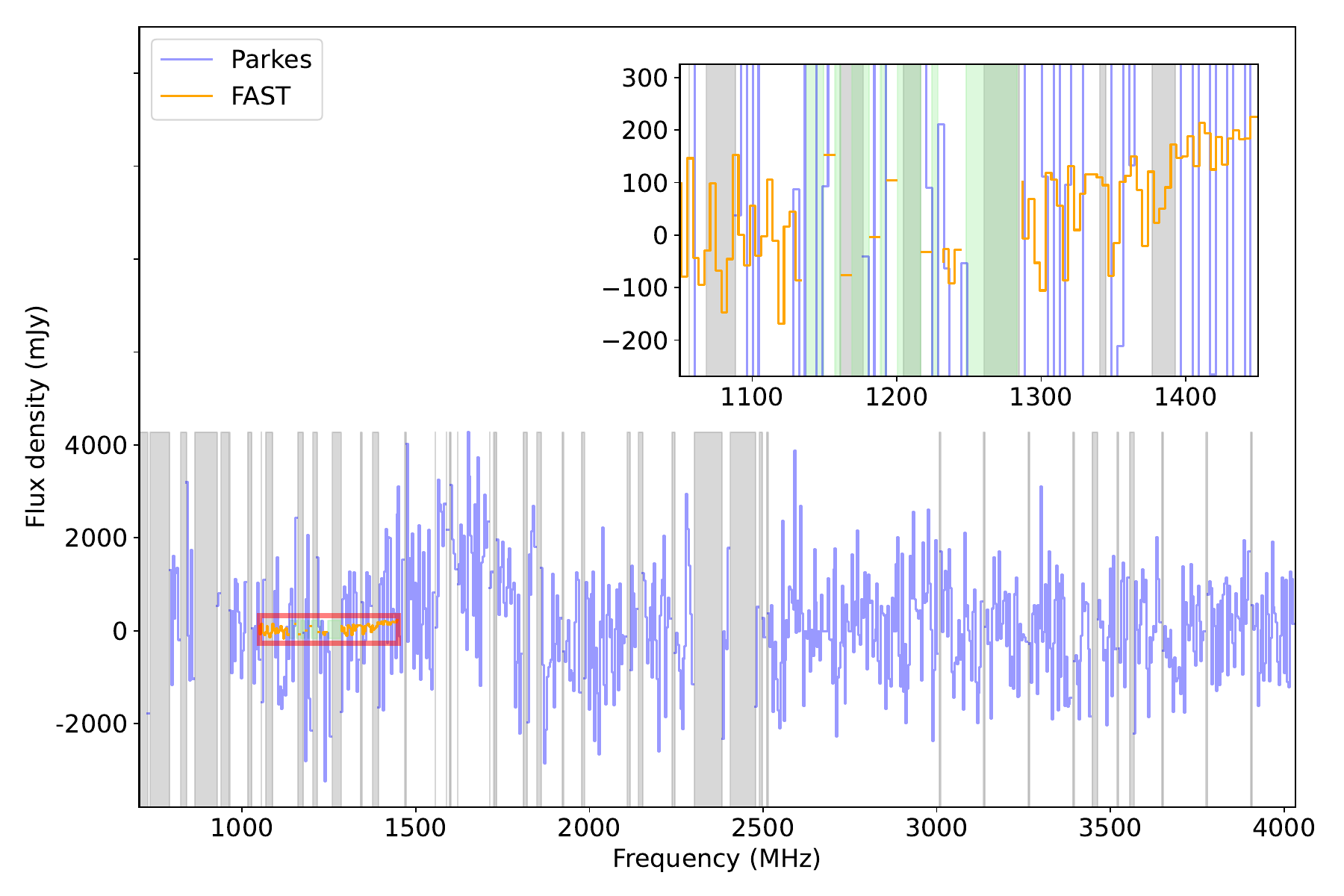}}
\vskip 2mm
\figcaption{7.5}{4}{
SEDs from Parkes and FAST.
The RFI channels in Parkes are indicated by the gray-shaded area, whereas the light green shading represents RFI in the FAST SED.
The zoomed-in SEDs in the frequency range of 1050-1450 MHz are shown in the upper right panel for better comparison.
Parkes shows significant fluctuations, obscuring the upward trend, while FAST reveals a clear band-limited emission and up-rising trend with lower fluctuations.
}
\medskip

Given that both SEDs originate from the same burst, and FAST's SED indicates a singular radiation component in the spectrum, the center frequency should align consistently between the two telescopes.
In our analysis of this burst, we initially determined the center frequency based on the SED fitting results from Parkes.
Subsequently, we use this center frequency to model the remaining parameters of FAST's SED using a Gaussian approach, thereby deriving FAST's model spectrum.
This method not only accurately captures the burst's radiation characteristics but also leverages FAST's enhanced sensitivity to precisely gauge changes in the lower frequency bands.

{\it Spectral sharpness.}
Narrowband and broadband radiation correspond to SEDs that vary differently in their sharpness with frequency.
In most pulsars, the SED of a pulse is typically modeled using a power-law function, with the spectral index indicating how the spectrum changes across different frequencies, thus constraining the radiation regime.
Currently, observed FRB bursts tend to exhibit narrowband radiation within a specific bandwidth range. 
Multiband detection reveals that a burst component generally exists only within a particular frequency range, differing from the single pulses of pulsars. 
Consequently, these bursts cannot be characterized by a spectral index using a power-law fit.
\vskip 4mm
\fl{5}\centerline{\includegraphics[width=\columnwidth]{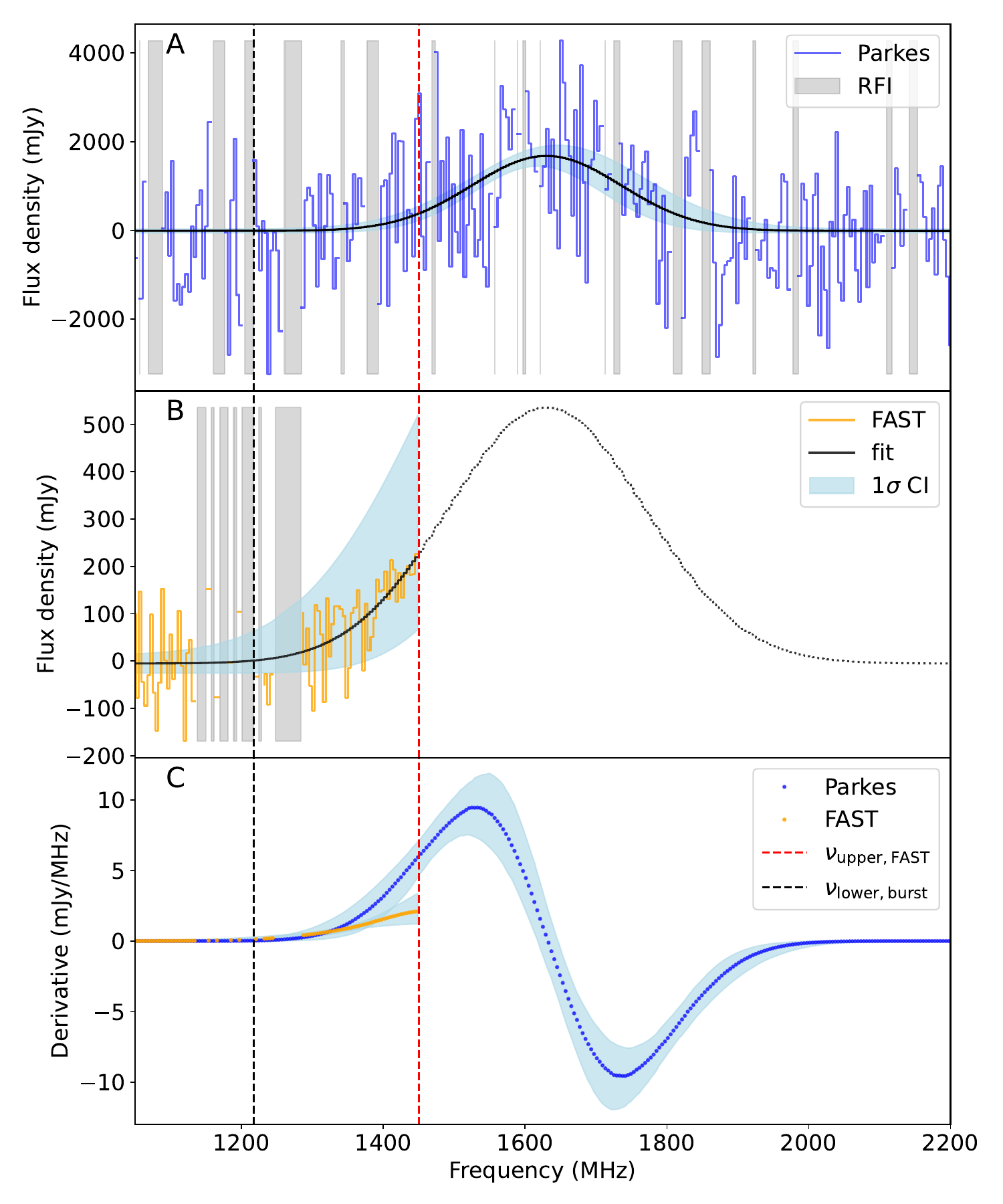}}
\vskip 2mm
\figcaption{7.5}{5}{
Panel A shows the SED and fitting result from Parkes, while Panel B shows results from FAST.
The light blue region represents the $1\sigma$ error range. 
The black dotted line in Panel B is the estimated SED beyond FAST's upper limit.
Panel C illustrates the SED derivatives over frequency. 
Summing these derivatives from the burst's lower frequency to FAST's upper limit allows us to compare spectral sharpness and determine the burst's radiation characteristics within the FAST frequency range.
}
\medskip

In previous FRB spectral studies, the ratio $\Delta \nu / \nu_c$ has been used to describe the spectral coverage of a burst, 
serving as a measure of the extent to which the spectral range of an FRB burst covers frequencies, thereby characterizing the narrowband feature of the burst.\ucite{38,39} 
Here, $\Delta \nu$ represents the bandwidth determined by fitting a Gaussian to the SED of the FRB burst, defined by the full width at half-maximum (FWHM), while $\nu_c$ is the central frequency of the burst.
This approach, although capable of describing bandwidth coverage, cannot capture the intensity of spectral changes from an energy perspective due to the lack of energy spectrum information.
Moreover, this method is only valid when bursts occur within the effective bandwidth; otherwise, instrumental effects may lead to erroneous statistical outcomes.
For instance, taking the burst simultaneously detected by Parkes and FAST, the burst observed by FAST approaches the upper edge of the frequency range, whereas the SED recorded by Parkes fully lies within its effective bandwidth.
The computed $\Delta \nu / \nu_c$ for each is shown in Table~\ref{tbl:2}.
While this method allows for comparison of spectral coverage between sources, it does not confirm whether a pulse has multiple radiation components in the spectrum.

The narrower bandwidth of FAST does not provide the correct peak frequency and bandwidth, rendering its values unable to accurately describe the burst's spectral nature. 
This bandwidth limitation compromises the originally reasonable results from Parkes.
We propose the use of the spectral sharpness method for analyzing all single bursts of FRBs.
By combining this with bandwidth distribution, we can statistically and verifiably characterize the single-pulse radiation of repeaters as narrowband emission from the SED. 
This approach allows for a more systematic analysis of the radiation mechanisms of FRBs and provides insight into their radiation characteristics from a spectral perspective.

We define spectral sharpness as the sum of the derivatives of the SED from the peak portion of the bandwidth to the lowest or highest frequency position of the visible burst,
\begin{equation}
    D = \sum_{i=1}^{N-1} \left( \frac{S(\nu_{i+1}) - S(\nu_i)}{\nu_{i+1} - \nu_i} \right),
    \label{eq:3}
\end{equation}
$D$ denotes the spectral sharpness, measured in units of $\mathrm{mJy/MHz}$, where $S$ represents the flux density of SED, $\nu$ is the frequency, and $i$ is the index of the frequency sample.
Since the peak frequency of the jointly detected burst exceeds FAST's detection limit, we can only analyze the sharpness of the SED in the low-frequency range. 
To reduce interference from insufficient integration time when calculating spectral sharpness, we first smoothed the SED and then performed Gaussian fitting. 
The fitted model spectrum was used to calculate their derivatives and spectral sharpness. 
\vskip 2mm
\tl{2}\tabtitle{7.8}{2}{
Bandwidth, frequency coverage, and spectral sharpness of the simultaneously detected burst.
``a'': Gaussian fit with fixed center frequency derived from the Parkes SED.
``b'': Direct full-parameter Gaussian fit to the SED of FAST.
``*'': Denotes $\Delta \nu$ cut within the FAST's bandwidth.
``$\dagger$'': Denotes $\Delta \nu$ given directly by Gaussian fitting without effective bandwidth cutting.
}
\vskip 2mm \tabcolsep 4.5pt
\centerline{\footnotesize
\begin{tabular}{lcccc}
\hline\hline\hline
& $\nu_{c}$ & $\Delta \nu$ & $\Delta \nu / \nu_c$ & $D$        \\
&   (MHz)   &    (MHz)     &                      &  (mJy/MHz) \\
\hline
$\mathrm{Parkes}$   & $1632.34_{-14.52}^{+14.01}$ & $ 248.1_{-33.3}^{+38.8} $     & $ 0.15_{-0.02}^{+0.02}$     & $97.45_{-41.79}^{+43.71}$ \\
$\mathrm{FAST}^{a}$ & $1632.34_{-14.52}^{+14.01}$ & $ {31.9_{-22.3}^{+28.0}}^* $  & $ {0.02_{-0.01}^{+0.02}}^* $ & $51.49_{-11.99}^{+12.34}$ \\
                    &                             & $ {326.1_{-60.5}^{+82.4}}^{\dagger}$ & ${0.20_{-0.04}^{+0.05}}^{\dagger}$ &              \\

$\mathrm{FAST}^{b}$ & $1541.80_{-65.16}^{+39.55}$   &${10.1_{-7.5}^{+13.3}}^*$           & ${0.01_{-0.00}^{+0.01}}^*$   & $47.84_{-10.04}^{+10.36}$ \\
                    &                               &${256.5_{-71.4}^{+69.5}}^{\dagger}$ & ${0.17_{-0.04}^{+0.04}}^{\dagger}$ &                     \\
\hline\hline\hline
\end{tabular}}
\medskip

If the SED from FAST exhibits a steeper change, it indicates a more rapid decrease in radiation at lower frequencies, suggesting a lack of additional emission components.
Figure~\ref{fig:5} and Table~\ref{tbl:2} show that the spectral sharpness values for FAST and Parkes are similar within the margin of error, though FAST's values are slightly lower. 
This discrepancy may be due to larger flux fluctuations in Parkes' data, which could affect the fitted bandwidth of its SED, resulting in higher calculated sharpness.
Moreover, the burst did not show weaker broadband radiation in the FAST band when comparing the SEDs from both telescopes. 
Combined with the observed changes in energy spectrum sharpness, we propose a narrowband radiation mechanism with a single emission component in the spectrum.

{\it Discussion.}
We proposed a method to compare the spectral sharpness to verify the characteristic of narrowband emission.
To illustrate our proposed method is reasonable, we compared our results to those of CHIME et al.,
who proposed using a ``running'' power-law (RPL) to model the frequency-dependent SED ($F_{k}$) from radio pulses.\ucite{40}
The ``running'' power-law function reads as follows,
\begin{equation}
    F_{k} = \left( \frac{\nu_{k}}{\nu_{r}} \right)^{\gamma + \beta\ln{(\nu_{k}/\nu_{r})}},
    \label{eq:4}
\end{equation}
where $F_{k}$ is the RPL SED function, $\nu_{k}$ is the electromagnetic frequency for channel k, $\nu_{r}$ is the reference (or ``pivot'') electromagnetic frequency, and $\kappa = \gamma + \beta \ln{(\nu / \nu_{r})}$ is the spectral index {depending} on $\nu$.

The distinction between broadband and Gaussian-like SEDs is in the values of $\left\{ \beta, \gamma \right\}$: broad signals have $\beta \approx 0$ and $\lvert \gamma \rvert \leqslant 10$, while Gaussian-like SEDs have large values for both parameters.\ucite{40}
The frequency coverage transition from a ``running'' power-law to Gaussian-like SEDs is characterized by
\begin{equation}
    \sigma^2 \approx-\frac{1}{2} \frac{\blue{\nu}_r^2}{\beta-\gamma / 2},
    \label{eq:5}
\end{equation}
where $\nu_{r}$ is the reference frequency, and $\sigma$ is the parameter in Gaussian function, which depicts the bandwidth of a spectrum.\ucite{40}
In the CHIME pipeline, the reference frequency was set to 400.1953125 MHz.
\vskip 4mm
\fl{6}\centerline{\includegraphics[width=0.85\columnwidth]{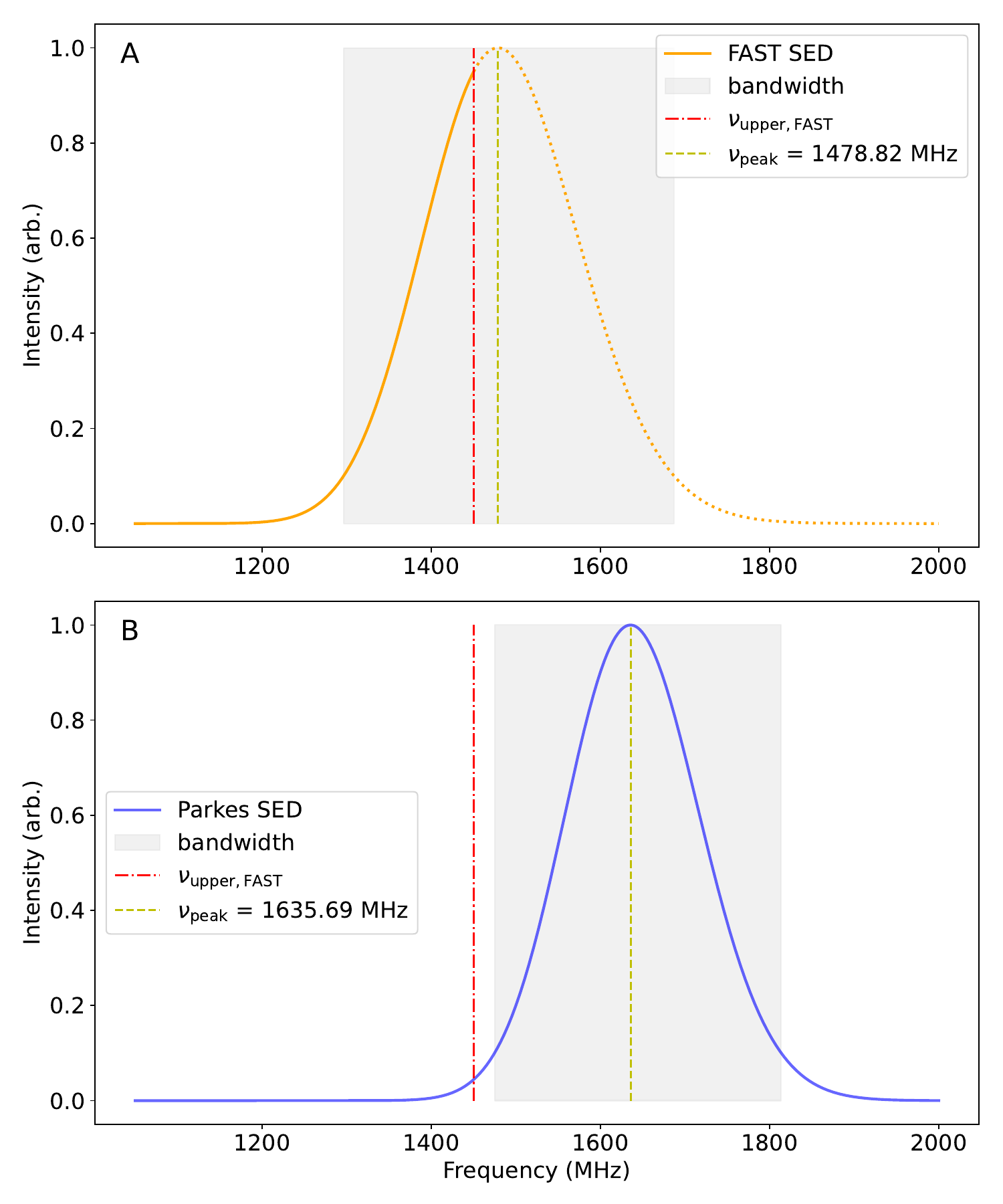}}
\vskip 2mm
\figcaption{7.5}{6}{
RPL SEDs from the CHIME fitting pipeline are normalized to [0, 1] for better visualization.
In the upper panel, the solid orange line shows the effective SED, with the dotted line indicating frequencies outside of FAST coverage.
The shaded region represents the bandwidths of FAST and Parkes.
The red dotted line marks FAST's upper limit at 1450 MHz, and the yellow dotted line shows the peak frequency for both telescopes.
}
\medskip

We modeled the single burst using CHIME's \software{fitburst.py}, setting the reference frequency to 1052 MHz for both FAST and Parkes. 
The derived $\beta_{\mathrm{FAST}}=90.40,\, \beta_{\mathrm{Parkes}}=189.80$, and $\gamma_{\mathrm{FAST}}=-132.70,\, \gamma_{\mathrm{Parkes}}=-215.06$, respectively.
The RPL indices suggest that the SEDs of both FAST and Parkes exhibit narrowband characteristics. 
According to Equation~(\ref{eq:5}), the SED detected by Parkes has a narrower bandwidth compared to that of FAST. 
Despite both telescopes displaying narrowband radiation features after SED fitting, FAST's limited bandwidth coverage did not fully capture the entire SED, as shown in Figure~\ref{fig:6}.

Consequently, the frequency discrepancy renders it impractical to compare results across different telescopes using the CHIME pipeline.
As a result, we adopted a traditional Gaussian fitting method utilizing the center frequency determined by Parkes as a fixed reference.
This approach facilitated the analysis of the burst radiation by combining spectral sharpness with an accurate bandwidth distribution.



Analysis of FAST data indicates that bursts from FRB 20190520B exhibit narrowband characteristics.
In observations, the narrow bandwidth of FRBs may result from intrinsic radiation mechanisms, propagation effects, and observational biases.\ucite{52,53,54}
Among propagation effects, plasma lensing and scintillation are two mechanisms that can narrow the FRB bandwidth.\ucite{52,54,57}
To avoid the influence of these effects on the bandwidth properties, we discussed the impact of these two effects on the bandwidth of the jointly observed burst.
In the SEDs observed by Parkes and FAST, no significant scintillation effects were detected at the common frequency of 1450 MHz, which both telescopes covered around this frequency range.
The flux densities at this frequency, recorded by both telescopes, were similar, indicating a continuous spectrum with narrowband features.
Given the higher sensitivity of FAST, which allowed it to capture a more complete SED bandwidth distribution at this frequency, it is unlikely that the narrowband radiation observed is a result of scintillation.
Therefore, we conclude that the burst detected by both telescopes is the same and its narrowband characteristics are primarily due to the intrinsic emission mechanism rather than to propagation effects.
Since both telescopes observed the same pulse, plasma lensing is also unlikely to have caused bandwidth narrowing.

Another mechanism potentially responsible for the observed narrowing of the burst bandwidth is the observational bias introduced by the flux threshold limitation.\ucite{52,54}
When the signal intensity of a burst falls below the telescope's detection threshold, parts of the signal that are below this threshold remain undetected, resulting in an observed bandwidth that is narrower than the actual signal.
This phenomenon leads to an observational bias towards a perceived narrower bandwidth.
This jointly observed burst displayed narrowband radiation characteristics in both the Parkes and FAST telescopes.
If the narrowband effect observed at Parkes results from insufficient sensitivity or a higher flux detection threshold, then using a higher-sensitivity telescope in joint observations can provide a bandwidth limit at a lower flux threshold.
FAST's sensitivity, approximately 22-27 times greater than Parkes over the same bandwidth range, enables FAST to detect radiation components with flux densities above 0.03788 Jy.
Despite this heightened sensitivity, the spectral energy distribution (SED) of the burst still exhibits narrowband characteristics at FAST, fitting well with a single Gaussian component.
This suggests that no additional radiation components were detectable within FAST's frequency range, confirming that the narrowband characteristic of this burst, as observed at FAST's detection threshold, stems from an intrinsic radiation mechanism.
To determine whether this feature is common across all repeaters, further observations are necessary.
Additionally, continued multiband studies are essential for evaluating theoretical models, such as Lyu et al.'s frequency comb model and the coherent curvature radiation beam model, which will further enrich our understanding of the radiation mechanisms of repeating bursts.\ucite{46,47,48}

{\it Summary and Conclusion.}
In this study, we conducted observations of FRB 20190520B using the FAST and Parkes telescopes, detecting approximately 300 bursts with each, which indicated similar event rates across their respective frequency ranges.
Our analysis confirmed the narrowband nature of a burst within the Parkes bandwidth, further supported by the higher sensitivity of FAST, which revealed that there were no additional radiation components in the spectrum.
Continued joint multiband observations are crucial to monitor the evolution of the dispersion measure (DM), analyze energy distributions, and improve our understanding of narrowband radiation processes, including periodicity, activity windows, and frequency dependencies, similar to those observed in FRB 20180916B.
Such efforts are vital not only for validating the radiation mechanisms and comprehending the propagation dynamics of FRBs, but also for investigating common origins among different types of repeating bursts.
This could significantly advance our understanding of their radiation models and underlying physical mechanisms.

In our analysis, we employed three methods to examine the spectral energy distribution (SED) and the bandwidth of the bursts. 
Although \software{fitburst.py} did not capture the complete SED of FAST and Parkes, it successfully identified a narrowband emission feature.
Additionally, the \(\Delta \nu / \nu_c\) method also revealed that the burst from FRB 20190520B exhibits the band-limited emission characteristic.
These methods collectively confirm the results obtained from our spectral sharpness approach, which effectively illustrates the energy changes and emission components in SEDs that were not discernible with previous methodologies. 

Our findings advocate for the use of SED and its sharpness as a novel approach to model the emission features of FRB bursts. 
This method not only confirmed the narrowband emission features of the burst observed simultaneously by FAST and Parkes, but also provided a framework for systematic comparison of bursts from different classes of FRBs.
Moving forward, we plan to extend this analysis to include a broader range of repeaters and compare these with bright single pulses and giant pulses from pulsars, thereby enriching our comparative studies of FRB radiation properties. 

\textit{Acknowledgements.}
This work was supported by National Science Foundation of China (No. 11988101, 12203069, 12041302) and the National SKA Program of China/2022SKA0130100. 
This work was also supported by the National Natural Science Foundation of China grant No. 12203045, the Office of the Leading Group for Cyberspace Affairs, CAS (No. CAS-WX2023PY-0102), and the CAS Youth Interdisciplinary Team and the Foundation of Guizhou Provincial Education Department for Grants No. KY(2023)059.
D.L. is a New Cornerstone investigator.
P.W. acknowledges support from the National Natural Science Foundation of China (NSFC) Programs No.11988101, 12041303, the CAS Youth Interdisciplinary Team, the Youth Innovation Promotion Association CAS (id. 2021055), and the Cultivation Project for FAST Scientific Payoff and Research Achievement of CAMS-CAS.
Y.F. is supported by National Natural Science Foundation of China grant No. 12203045, by the Leading Innovation and Entrepreneurship Team of Zhejiang Province of China grant No. 2023R01008, and by Key R\&D Program of Zhejiang grant No. 2024SSYS0012. 
Xianghan Cui is supported by the China Scholarship Council (Grant No. 202304910441).

\end{multicols}
\end{document}